\documentclass[pdf,twocolumn,superscriptaddress,preprintnumbers,amsmath,amssymb]{revtex4}
\usepackage{mathrsfs}
\usepackage{float,epsfig}
\usepackage{graphicx}% Include figure files
\usepackage{dcolumn}% Align table columns on decimal point
\usepackage{bm}% bold math
\usepackage{amsmath,amssymb,epsfig,float,graphics}
\usepackage{amsmath,amssymb,amsthm}
\usepackage[colorlinks=true,linkcolor=blue]{hyperref}
\usepackage{subfigure}
\usepackage{booktabs}
\usepackage[mathscr]{euscript}

\usepackage{graphicx}
\usepackage{epstopdf}
\newcommand{\bea}{\begin{eqnarray}}
\newcommand{\eea}{\end{eqnarray}}
\newcommand{\beq}{\begin{equation}}
\newcommand{\eeq}{\end{equation}}
\newcommand{\nn}{\nonumber}
\def\/{\over}

\usepackage{color}

\DeclareMathOperator{\csch}{csch}
\begin{document}
\baselineskip=0.45 cm

\title{Probing cosmic string spacetime through parameter estimation}

\author{Ying Yang\footnote{Corresponding author, Email: yingyanghnust@163.com}}
\affiliation{Department of Physics, Key Laboratory of Intelligent Sensors and Advanced Sensor Materials, Hunan University of Science and Technology, Xiangtan, Hunan 411201, P. R. China.}

\author{Jiliang Jing}
\affiliation{Department of Physics, Key Laboratory of Low Dimensional Quantum Structures and Quantum Control of Ministry of Education, and Synergetic Innovation Center for Quantum Effects and Applications, Hunan Normal University, Changsha, Hunan 410081, P. R. China.}

\author{Zehua Tian\footnote{Corresponding author, Email: tianzh@ustc.edu.cn}}
\affiliation{CAS Key Laboratory of Microscale Magnetic Resonance and Department of Modern Physics, University of Science and Technology of China, Hefei 230026, China}
\begin{abstract}
Quantum metrology studies the ultimate precision limit of physical quantities by using quantum strategy. In this paper we apply the quantum metrology technologies to the relativistic framework for estimating the deficit angle parameter of cosmic string spacetime. We use a two-level atom coupled to electromagnetic fields as the probe and derive its dynamical evolution by treating it as an open quantum system. We estimate the deficit angle parameter by calculating its quantum Fisher information(QFI). It is found that the quantum Fisher information depends on the deficit angle, evolution time, detector initial state, polarization direction, and its position. We then identify the optimal estimation strategies, i.e., maximize the quantum Fisher information via all the associated parameters, and therefore optimize the precision of estimation. Our results show that for different polarization cases the QFIs have different behaviors and different orders of magnitude, which may shed light on the exploration of cosmic string spacetime.
\end{abstract}

\baselineskip=0.45 cm
\maketitle
\newpage

\section{Introduction}

In the context of quantum field theory, cosmic strings have attracted considerable attention due to their significance on the research of astrophysical, gravitational and cosmological~\cite{Kaiser,Bennett,Vilenkin,Hindmarsh}. Although the cosmic microwave background radiation shows that the cosmic strings are not abundant, the evolution of string brings distinct astrophysical effects, such as gammaray bursts~\cite{Brandenberger1,MacGibbon}, producing detectable gravitational waves~\cite{Damour,Brandenberger2,Jackson}, and high-energy cosmic rays~\cite{Cheng}. Among these interesting astrophysical consequences caused by cosmic string, topology defect plays an important role~\cite{Kibble}. The interest to the research of topology defects in cosmic string spacetime is amplified by various theoretical studies, which have created a promising perspective for the development of this field~\cite{Vilenkin,Hindmarsh,Davis1,Siemens}.

Based on the unique signatures of cosmic string~\cite{Linet,Frolov,Davies1,Gott,Charnock,Foreman}, various investigations have been conducted to detect the string~\cite{Dvorkin, Eiichiro,Aghanim,Urrestilla,Helliwell}. The simplest cosmic string spacetime is characterized by a flat metric with a deficit angle, which is described by an infinite, straight and static cylindrically symmetric cosmic string, and many quantum effects exhibit significant characteristics in such spacetime~\cite{Sousa,Bezerra,Iliadakis,Berezinsky,Davies,Bilge}.

Quantum metrology which aims to achieve higher precision than the classical ways by using quantum techniques, has been paid a lot of attention recently~\cite{Giovannetti,Giovanetti2,Paris,advance}. There are considerable interests in applying quantum metrological techniques in relativistic settings, specifically, to relativistic quantum fields \cite{Ahmadi,Huang,Kish,Mann R B, xiaobao}. One of the purpose is to explore, in the relativistic quantum regimes, how the relativistic effects on quantum system affect the precision of certain measurements~\cite{Parameter estimation,relativistic motion, Liu1, M-Ahmadi, Liu2}. The other is how to measure the physical quantity of relativity, e.g., Unruh-Hawking temperature, with quantum techniques to obtain higher precision beyond the classical case~\cite{Tian, Adesso, TianZH}. Previous research has investigated the metrology of a wide range of relativistic phenomena, including measuring the distance to achieve adequate sensitivity for the detection of gravitational wave~\cite{gravitational wave}, high precision estimation of the Earth's Schwarzschild spacetime parameters by quantum experiments~\cite{Schwarzschild spacetime parameters}, estimation of gravitational redshift from matter wave interference~\cite{gravitational redshift}, and so on.

Since in practice a lot of quantities of interest to us do not correspond to quantum observable, direct observation of them is not accessible. In these situations, one has to infer the interested values of the quantities by examining a set of data from the measurement of a different observable, or a set of observable. Therefore, any conceivable strategy aimed at evaluating the quantity of interest eventually reduces to a parameter-estimation problem~\cite{Helstrom,Holevo}. In this regard, let us note that the quantum Fisher information (QFI) is the key figure of merit in the framework of quantum estimation theory. The QFI links to the Cram{\'{e}}r-Rao bound and determines the ultimate bound on the precision of the estimator~\cite{Helstrom}. Larger QFI means more optimal precision for estimating parameter can be possibly achieved in a metrological task. Therefore, how to maximize the QFI in a certain system via different system parameters and different approaches is particularly significant.

We now study quantum metrology in the relativistic framework and focus on estimating the deficit angle in cosmic string spacetime. We use a two-level atom as probe which is coupled to electromagnetic fields in the background of cosmic string spacetime. We study the dependence of the QFI on various parameters, and identify strategies for maximizing it. Note that although relativistic metrology has been considered in the cosmic string spacetime, all of them~\cite{jin, huang} paid their attention to how the spacetime property affects on estimating the quantum state parameter of probe, and overlooked the estimation of the spacetime parameter, e.g., the deficit angle. However, this overlooked issue plays a very important role in the exploration of cosmic string physics~\cite{conicity}, since it might tell us how to
obtain the ultimate limit of precision for estimating cosmic string
and what the ultimate precision bounded by quantum mechanics is. In our previous research~\cite{ying}, we estimated the deficit angle of a cosmic string spacetime using a moving detector that is coupled to a massless scalar field at vacuum state. To further elaborate the parameter estimation of cosmic string spacetime, we here extent the relevant investigation to a more practical scenario by replacing the massless scalar field with
a vector field, i.e., an electromagnetic field. We would like to emphasize that compared with the massless scalar field case,
the unique emerging electromagnetic polarization will cause significant effects on the QFI for estimating the deficit angle. Specifically,
the QFIs not only have different behaviors via dependent parameters, but also there is more than two orders of QFI magnitude
difference for different polarization cases, as shown in the following.

The outline of this paper is as follows. In Sec. \ref{section2}, we recall the physics of QFI. Then we introduce our detector, a two-level atom which is coupled to electromagnetic fields in the cosmic string spacetime in Sec. \ref{section3}. In Sec. \ref{section4} we investigate the parameter estimation of the deficit angle parameter with different polarizations by using a static atomic detector, and find the effective strategies to maximize the QFI for estimating the deficit angle. In Sec. \ref{section5}, we give a simple discussion about our model. Final remarks and conclusions are given in Sec. \ref{section6}.

%%%%%%%%%%%%%%%%%%%%%%%%%%%%%%%%%%%%%%%%%%%%%%%%%
\section{Quantum Fisher information} \label{section2}

The quantum Cram\'{e}r-Rao bound provides a fundamental limit on the precision that quantum measurements can achieve~\cite{Cramer,Helstrom,Holevo}, and it is expressed as
\begin{eqnarray}\label{QFI}
{\rm Var}(\lambda)\geq\frac{1}{N F_\lambda}.
\end{eqnarray}
Here $\rm{Var}(\lambda)=E_{\lambda}[(\hat{\lambda}-\lambda)^2]$ is the variance of the estimator, $N$ is the number of measurements and $F_\lambda=F(\lambda)$ denotes the QFI of parameter $\lambda$. The definition of $F(\lambda)$ can be written as~\cite{Cramer}
\begin{equation}
F(\lambda)\equiv Tr[\rho(\lambda)L(\lambda)^2]\;,
\end{equation}
where $\rho$ is the density matrix of the detector and $L(\lambda)$ is the symmetric logarithmic derivative which satisfies $\partial \rho(\lambda)=\frac{1}{2}(L(\lambda)\rho(\lambda) + \rho(\lambda)L(\lambda))$. The quantum Cram\'{e}r-Rao bound provides the ultimate bound of parameter estimation accuracy for a state of the family $\rho(\lambda)$. Consider a two-level quantum system, whose quantum state in the Bloch sphere is of the form
\begin{equation}\label{Bloch}
\rho=\frac{1}{2}(I+\vec{\omega}\cdot\vec{\sigma})\;,
\end{equation}
where $\vec{\omega}=(\omega_1,\omega_2,\omega_3)$ denotes the Bloch vector, and $\vec{\sigma}=(\sigma_1,\sigma_2,\sigma_3)$ are the Pauli matrices.
For such an elemental quantum system, one can find its QFI expression analytically~\cite{Zhong}
\begin{equation}\label{F}
F_{\lambda}=\left\{
    \begin{array}{l}
 |\partial_{\lambda}\vec{\omega}|^2+\frac{(\vec{\omega}\cdot\partial_{\lambda}\vec{\omega})^2}{1-|\vec{\omega}|^2},\;\;\;|\vec{\omega}|<1\;,  \\
    |\partial_{\lambda}\vec{\omega}|^2,\;\;\;\;\;\;\;\;\;\;\;\;\;\;\;\;\;\;\;\;|\vec{\omega}|=1. \\
    \end{array}
    \right.
\end{equation}

%%%%%%%%%%%%%%%%%%%%%%%%%%%%%%%%%%%%%%%%%%%%%%%
\section{Open quantum system in cosmic string spacetime}\label{section3}

In practice, any quantum systems inevitably interact with external environment, which may lead to dissipation and decoherence. Correspondingly the ambient noise will have effect on the QFI encoded in the system's quantum state and thus have a certain influence on the accuracy of parameter estimation. Then, it is very necessary to explore the parameter estimation issues in the area of open quantum system. In this section, we will introduce a two-level detector
which is subjected to the quantum vacuum fluctuations of an electromagnetic field in the cosmic string spacetime. The detector is treated as an open quantum system and its dynamics is derived by tracing over the degree of freedom of the electromagnetic field.

Let us begin with the cosmic string spacetime, whose metric is of the form~\cite{Linet,Skarzhinsky,jin}
\begin{equation}\label{1}
ds^2=dt^2-dr^2-r^2d{\alpha}^2-dz^2.
\end{equation}
As shown in Fig.~\ref{fig1}, in the cylindrical coordinates the cosmic string spacetime is supposed to have an string lies along the $z$ direction.
\begin{figure}
\begin{center}
\includegraphics[scale=0.8]{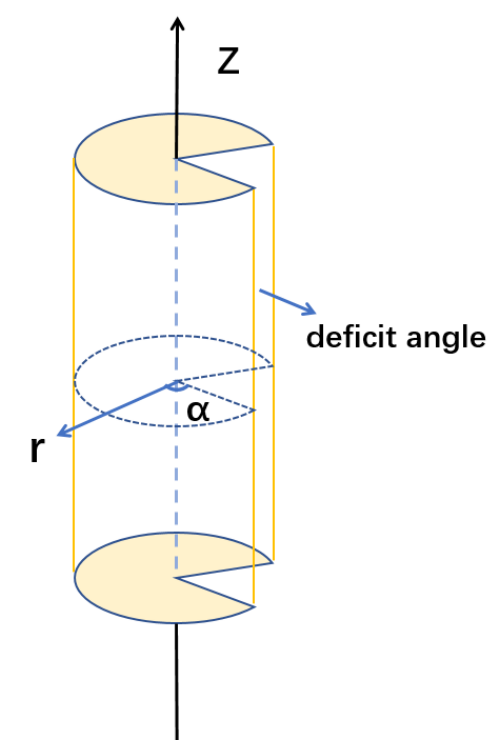}
\caption{ Schematic representation of the cosmic string spacetime with an infinite and straight cosmic string along $z$ direction.}
\end{center}
\end{figure}\label{fig1}
The metric \eqref{1} is quite similar to the metric of Minkowski spacetime, the difference is that for the Minkowski spacetime $0\leq \alpha < 2\pi$, while for the cosmic string spacetime $0\leq \alpha < \frac{2\pi}{\nu}$, where $\nu=(1-4G\mu)^{-1}$ with $G$, the Newton constant, and $\mu$, the mass per unit length of the string. Therefore, compared to the Minkowski spacetime, the cosmic string spacetime is locally flat but with a deficit angle which is used to characterize the nontrivial global topology. One can quantize the electromagnetic field in the cosmic string spacetime, the corresponding vector potential of electromagnetic field is expressed as~\cite{Pavel}
\beq
A_{\xi}(t,\vec{x})=\int d \mu_j\;[\;c_{\xi j}(t)f_{\xi j}(\vec{x})+c_{-\xi j}^{\dag}(t)f_{-\xi j}^{\ast}(\vec{x})]\label{Axi}
\eeq
with
\beq
\int d \mu_j=\sum _{m=-\infty}^{\infty}\int_{-\infty}^{\infty}d k_{3}\int_{0}^{\infty}d k_{\perp}k_{\perp}\;,
\eeq
in which $c_{\xi j}(t)=c_{\xi j}(0)e^{-i\omega t}$ is the annihilation operator and $c^{\dag}_{-\xi j}(t)=c^{\dag}_{-\xi j}(0)e^{i\omega t}$ is the creation operator with $\xi\in\{0,\pm1,3\}$, $m\in Z$, $k_{3}\in[-\infty,\infty]$, $k_{\perp}\in[0,\infty]$. The exact form of the $f_{\xi j}(\vec{x})$ is $f_{\xi j}(\vec{x})=\frac{1}{2\pi}\sqrt{\frac{\nu}{2\omega}}J_{|\nu m+\xi|}(\kappa_{\bot}r)e^{i(\nu m\alpha+k_3z)}$.  It is shown that
\bea
c_{\xi j}(0)&=&i\int d^3\vec{x} f^{\ast}_{\xi j}(t,x)\overleftrightarrow{\partial_{t}}A_{\xi}(t,x)\;,\\
c^{\dag}_{\xi j}(0)&=&-i\int d^3\vec{x} f_{\xi j}(t,x)\overleftrightarrow{\partial_{t}}A_{\xi}(t,x)\;,
\eea
where the commutation relations are as follows
\bea
&&[c_{\xi j}(t),c_{\xi j'}^{\dag}(t)]=\delta_{j,j'}\quad\; for \quad\xi=\pm 1,3,\\
&&[c_{0 j}(t),c_{0 j'}^{\dag}(t)]=-\delta_{j,j'} \;\;
 for \quad \xi=0.
\eea
For more details of the quantization of the electromagnetic field in the cosmic string spacetime, one can refer to Ref.~\cite{Pavel}.

We model a two-level atom as the detector which is coupled to an electromagnetic field in the cosmic string spacetime. The total Hamiltonian of the whole system is
\begin{eqnarray}
H=H_{A}+H_{F}+H_{I}\;,
\end{eqnarray}
where $H_{A}$, $H_{F}$ and $H_{I}$ are the atomic Hamiltonian, the field Hamiltonian and the interaction Hamiltonian, respectively. The specific expression of the atomic Hamiltonian is $H_{A}=\frac{1}{2}\hbar\omega_{0}\sigma_{3}$, where $\hbar\omega_{0}$ is the energy spacing of the atom and $\sigma_3$ represents the Pauli matrix. The interaction Hamiltonian is described as
$H_{I}=-e\texttt{r}\cdot\texttt{E}(x(\tau))$, where $e$ is electric charge, $e\texttt{r}$ denotes the atomic electric dipole moment, and $\texttt{E}(x(\tau))$ represents the strength of the electric field along the atom's trajectory $x(\tau)$.

We assume the total density matrix of the detector-field system at the beginning can be written as $\hat{\rho}_{\rm tot}(0)=\hat{\rho}(0)\otimes|0\rangle\langle0|$, with $\hat{\rho}(0)$ being the atomic initial density matrix, and $|0\rangle\langle0|$ representing the vacuum state of the field. In the atomic frame, the entire quantum system follows the von Neumann equation~\cite{Heinz}
\begin{eqnarray}\label{von Neumann equation}
\frac{\partial\hat{\rho}_{\rm tot}(\tau)}{\partial\tau}&=&-\frac{i}{\hbar}[H,\hat{\rho}_{\rm tot}(\tau)]\;,
\end{eqnarray}
where $\tau$ is the proper time. After tracing over the field degree of freedom $\hat{\rho}(\tau)=Tr_{F}[\hat{\rho}_{\rm tot}(\tau)]$, the master equation of the atom is given under the Born-Markov approximation as~\cite{Lindblad}
\begin{eqnarray}\label{Lindblad equation}
\frac{\partial\hat{\rho}(\tau)}{\partial\tau}&=&-\frac{i}{\hbar}[H_{eff},\hat{\rho}(\tau)]+\mathcal{L}[\hat{\rho}(\tau)]\;,
\end{eqnarray}
here ${\cal L}[\rho]={1\over2} \sum_{i,j=1}^3
a_{ij}\big[2\,\sigma_j\rho\,\sigma_i-\sigma_i\sigma_j\, \rho
-\rho\,\sigma_i\sigma_j\big]$ with $a_{ij}=A\delta_{ij}-iB\epsilon_{ijk}\delta_{k3}-A\delta_{i3}\delta_{j3}$, and we have used the definition $A=\frac{1}{4}[{\cal {G}}(\omega_0)+{\cal{G}}(-\omega_0)]$ and $B=\frac{1}{4}[{\cal {G}}(\omega_0)-{\cal{G}}(-\omega_0)]$.
By defining the two-point functions of the electromagnetic field
\begin{equation}
G^{+}(x-x')={e^2\/\hbar^2} \sum_{i,j=1}^3\langle +|r_i|-\rangle\langle -|r_j|+\rangle\,\langle0|E_i(x)E_j(x')|0 \rangle\;,
\end{equation}
with $\langle0|E_i(x)E_j(x')|0\rangle=\partial_i\partial'_j\langle0|A_0(x)A_0(x')|0\rangle+
\partial_0\partial'_0\langle0|A_i(x)A_j(x')|0\rangle$, we achieve its Fourier and Hilbert transforms as follows
\begin{equation}
{\cal G}(\lambda)=\int_{-\infty}^{\infty} d\Delta\tau \,
e^{i{\lambda}\Delta\tau}\, G^{+}\big(\Delta\tau\big)\; ,
\end{equation}
and
\begin{equation}
{\cal K}(\lambda)=\frac{P}{\pi
i}\int_{-\infty}^{\infty} d\omega\ \frac{ {\cal G}(\omega)
}{\omega-\lambda} \;.
\end{equation}
Then the effective Hamiltonian $H_{\rm eff}$ is
\begin{equation}\label{heff}
H_{\rm eff}=\frac{1}{2}\hbar\Omega\sigma_3={\hbar\over 2}\{\omega_0+{i\/2}[{\cal
K}(-\omega_0)-{\cal K}(\omega_0)]\}\,\sigma_3\;,
\end{equation}
where $\Omega$ is the effective level spacing of the atom, containing the free term $\omega_0$ and the Lamb shift expressed as
the second term on the right of Eq. \eqref{heff}. We assume that the detector is initially prepared in an arbitrary state $|\psi(0)\rangle=\cos\frac{\theta}{2}|+\rangle+e^{i\phi}\sin\frac{\theta}{2}|-\rangle$, where $\theta$ and $\phi$ denote the weight parameter and phase parameter, and $|-\rangle$, $|+\rangle$ represent the ground state and excited state of the atom, respectively. With this initial condition, by solving Eq. \eqref{Lindblad equation} we find the time-dependent Bloch vector of the detector as~\cite{Kimura}
\begin{align}\label{omega}
&\omega_1(\tau)=\sin\theta \cos(\Omega\tau+\phi)e^{-\frac{1}{2}A\tau}\;,\nonumber\\
&\omega_2(\tau)=\sin\theta \sin(\Omega\tau+\phi)e^{-\frac{1}{2}A\tau}\;,\nonumber\\
&\omega_3(\tau)=\cos\theta e^{-A\tau}-\frac{B}{A}(1-e^{-A\tau})\;.
\end{align}
The interaction between the detector and field is encoded into the parameters, $A$ and $B$, which determines the detector's evolution and contains the information of spacetime due to scattering off the quantum field. Therefore, one can infer the related parameters of spacetime (what we will estimate in the
following) through the relevant outcomes by performing the measurements on the detector.

%%%%%%%%%%%%%%%%%%%%%%%%%%%%%%%%%%%%%%%%%%%%%
\section{Parameter estimation in cosmic string spacetime}\label{section4}

In this section, we intend to exploit local quantum estimation theory to find the ultimate limits of precision of the deficit angle parameter in the detector-field model. Specifically, we study how the QFI is affected by the detector's state parameters, such as evolution time, detector initial state, polarization direction, and so on.

\subsection{Static detector in cosmic string spacetime}
We now probe the cosmic string spacetime by using a static detector which is coupled to a vacuum fluctuating electromagnetic field. The trajectory of the detector is\cite{jin}
\begin{align}\label{traj1}
t(\tau)=\tau\;,~~~~~
r(\tau)=r\;,\\\nonumber
\alpha(\tau)=\alpha\;,~~~~~
z(\tau)=z\;,
\end{align}
which is described in the polar coordinates. Substituting the above trajectory \eqref{traj1} into the Wightman function of the electromagnetic field, we find its final form along the detector's trajectory can be written as
\begin{widetext}
\begin{eqnarray}\label{green1}
\langle0|E_{r}(x)E_{r}(x')|0\rangle&=&\frac{\nu}{8\pi^{2}}\int d\mu_j e^{i\omega(\tau-\tau')}
\biggl[\frac{\omega}{2}(J_{|\nu m+1|}^{2}(k_{\perp}r)+J_{|\nu m-1|}^{2}(k_{\perp}r))-\frac{1}{\omega}\biggl(\frac{d J_{|\nu m|}(k_{\perp}r)}{d r}\biggr)^2\biggr]\;,\\
\langle0|E_{\alpha}(x)E_{\alpha}(x')|0\rangle&=&\frac{\nu r^2}{8\pi^{2}}\int d\mu_j e^{i\omega(\tau-\tau')} \biggl[\frac{\omega}{2}(J_{|\nu m+1|}^{2}(k_{\perp}r)+J_{|\nu m-1|}^{2}(k_{\perp}r))-\frac{\nu^2 m^2}{\omega r^2}J_{|\nu m|}^2(k_{\perp}r)\biggr]\;,\\
\langle0|E_{z}(x)E_{z}(x')|0\rangle&=&\frac{\nu}{8\pi^{2}}\int d\mu_j e^{i\omega(\tau-\tau')}\frac{k_{\perp}^2}{\omega}J_{|\nu m|}^2(k_{\perp}r)\;.
\end{eqnarray}
\end{widetext}
Then the corresponding Fourier transform is given by
\begin{equation}
{\cal G}(\lambda)=\sum_i{e^2|\langle -|r_i|+\rangle|^2\lambda^3\/3\pi}f_{i}(\lambda,r,\nu)\Theta(\lambda),
\end{equation}
where $\Theta(\lambda)$ is the step function and
\begin{widetext}
\begin{eqnarray}
\nonumber
f_{r}(\omega,r,\nu)&=&\frac{3\nu}{4}\sum_m\int_0^{1}d\eta\frac{\eta}{\sqrt{1-\eta^2}}\bigg[(2-\eta^2)
J^2_{|\nu m+1|}(\omega r \eta)+\eta^2J_{|\nu m|-1}(\omega r \eta)J_{|\nu m|+1}(\omega r \eta)\bigg]\;,\label{definition f1}\nn
\\
\nonumber
f_{\alpha}(\omega,r,\nu)&=&\frac{3\nu}{4}\sum_m\int_0^{1}d\eta\frac{\eta}{\sqrt{1-\eta^2}}\bigg[(2-\eta^2)
J^2_{|\nu m+1|}(\omega r \eta)-\eta^2J_{|\nu m|-1}(\omega r \eta)J_{|\nu m|+1}(\omega r \eta)\bigg]\;,\label{definition f2}\nn
\\
f_{z}(\omega,r,\nu)&=&\frac{3\nu}{2}\sum_m\int_0^{1}d\eta\frac{\eta^3}{\sqrt{1-\eta^2}}J^2_{|\nu m|}(\omega r \eta)\;.\label{definition f3}
\end{eqnarray}
\end{widetext}

\iffalse
\begin{eqnarray}
\nonumber
f_{r}(\omega,r,\nu)&=&\frac{3\nu}{4}\sum_m\int_0^{1}d\eta\frac{\eta}{\sqrt{1-\eta^2}}\bigg[\frac{(2-\eta^2)}
{J^{-2}_{|\nu m+1|}(\omega r \eta)}
\\
\nonumber
&&+\frac{\eta^2J_{|\nu m|-1}(\omega r \eta)}{J^{-1}_{|\nu m|+1}(\omega r \eta)}\bigg]\;,\label{definition f1}\nn
\\
\nonumber
f_{\alpha}(\omega,r,\nu)&=&\frac{3\nu}{4}\sum_m\int_0^{1}d\eta\frac{\eta}{\sqrt{1-\eta^2}}\bigg[\frac{(2-\eta^2)}
{J^{-2}_{|\nu m+1|}(\omega r \eta)}
\\
\nonumber
&&-\frac{\eta^2J_{|\nu m|-1}(\omega r \eta)}{J^{-1}_{|\nu m|+1}(\omega r \eta)}\bigg]\;,\label{definition f1}\nn
\\
f_{z}(\omega,r,\nu)&=&\frac{3\nu}{2}\sum_m\int_0^{1}d\eta\frac{\eta^3}{\sqrt{1-\eta^2}}J^2_{|\nu m|}(\omega r \eta)\;.\label{definition f3}
\end{eqnarray}
\fi

\begin{figure}
\begin{center}
\includegraphics[scale=0.9]{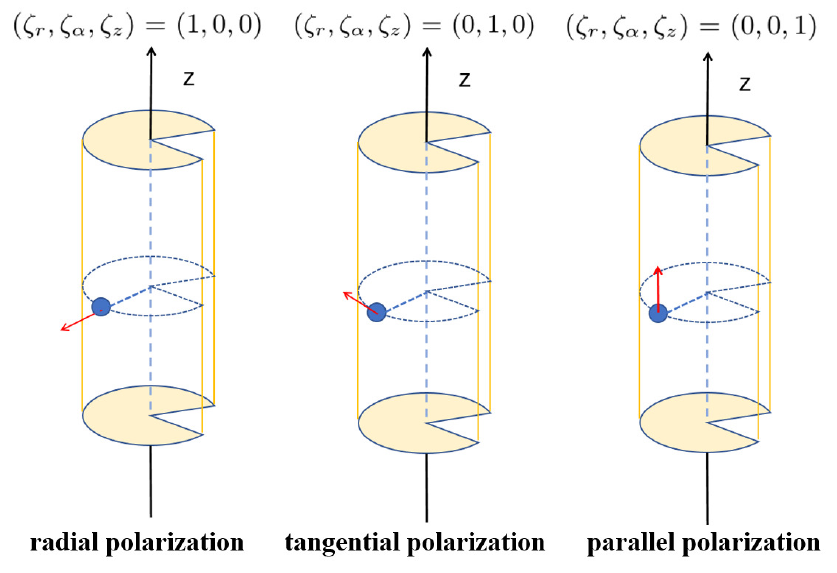}
\caption{Polarization schematic: the left panel to the right panel represent the radial polarization, the tangential polarization and the parallel polarization, respectively.}\label{fig2}
\end{center}
\end{figure}
With the above Fourier transform, we can find the coefficients of the Kossakowski matrix $a_{ij}$ is given by
\beq\label{A-B}
A=B={\gamma_0\/4}\sum_i\zeta_{i}f_{i}(\lambda,r,\nu),
\eeq
where $\gamma_0=e^2|\langle -|{\bf r}|+\rangle|^2\,\omega_0^3/3\pi$, and $\zeta_{i}=|\langle -|r_i|+\rangle|^2/|\langle -|{\bf r}|+\rangle|^2$ represents the relative polarizability which satisfies $\sum_i\zeta_{i}=1$, as shown in Fig. \ref{fig2}. We will adopt the notation $f(\lambda,r,\nu)=\sum_i\zeta_{i}f_{i}(\lambda,r,\nu)$ for simplification in the following. From Eqs. \eqref{F}, \eqref{omega}, and \eqref{A-B}, we can find the corresponding QFI for estimating the deficit angel parameter, $\nu$, is given by
\iffalse
\begin{widetext}
\begin{equation}
F_{\nu}(\omega,\nu,\tau,\theta, r)=\frac{e^{-f(\lambda,r,\nu)\tau}(\partial_{\nu}f(\lambda,r,\nu))^2\tau^2(2e^{f(\lambda,r,\nu)\tau}-1+\cos\theta)\cos^2\frac{\theta}{2}}
{2(e^{f(\lambda,r,\nu)\tau}-1)}\;,
\end{equation}
\end{widetext}
\fi
\begin{eqnarray}\label{F2}
\nonumber
F_{\nu}(\omega,\nu,\tau,\theta, r)&=&\frac{e^{-f(\lambda,r,\nu)\gamma_0\tau}(\partial_{\nu}f(\lambda,r,\nu))^2(\gamma_0\tau)^2\cos^2\frac{\theta}{2}}
{2(e^{f(\lambda,r,\nu)\gamma_0\tau}-1)}\\
&&\times
(2e^{f(\lambda,r,\nu)\gamma_0\tau}-1+\cos\theta).
\end{eqnarray}
For convenience, we adopt the notations $f_{r}$, $f_{\alpha}$ and $f_{z}$ as $f_{r}(\omega_0,r,\nu)$, $f_{\alpha}(\omega_0,r,\nu)$ and $f_{z}(\omega_0,r,\nu)$, and adopt the notation $F_{\nu}$ as $F_{\nu}(\omega,\nu,\tau,\theta, r)$ in the following. Note that different polarizations have different contributions to the parameters $A$ and $B$ discussed above, thus different effects on the QFI. Finding out what kind of polarization is optimal for the QFI should be a significant issue for the detection of cosmic string. We will investigate the QFI of deficit angel parameter $\nu$ with the radial polarization, the tangential polarization and the parallel polarization in the following.

\subsection{The estimation of deficit angel parameter for different polarizations}

For the radial polarization case, we have $(\zeta_r, \zeta_\alpha, \zeta_z)=(1, 0, 0)$ and the corresponding QFI is given by
\begin{equation}
F_{\nu}=\frac{e^{-f_{r}\gamma_0\tau}(\partial_{\nu}f_{r})^2(\gamma_0\tau)^2(2e^{f_{r}\gamma_0\tau}-1+\cos\theta)\cos^2\frac{\theta}{2}}
{2(e^{f_{r}\gamma_0\tau}-1)}\;.
\end{equation}
The QFI is independent on the  phase $\phi$ of the initial state, and it only depends on the evolution time, $\tau$, the deficit angel, $\nu$, the initial state parameter, $\theta$, and distance of the detector relative to the string, $r$. Here $r$ is in the unit of $\frac{c}{\omega_0}$, and $\tau$ is in the unit of $\frac{1}{\gamma_0}$. The units of $F_{\nu}$ depends on the deficit angle parameter. Since the deficit angle parameter is dimensionless, we obtain $F_{\nu}$ is dimensionless here. In order to ensure that the coordinates quantity in the following figures are dimensionless, we have worked with quantities by rescaling the time and the distance
\begin{eqnarray}\label{tau1}
\tau\longmapsto\widetilde{\tau}\equiv\gamma_0\tau,~~~~r\longmapsto\widetilde{r}\equiv r\omega_0.
\end{eqnarray}
For convenience, we continue to term $\widetilde{\tau}$ and $\widetilde{r}$ as $\tau$ and $r$, respectively in the following.

For the tangential polarization $(\zeta_r, \zeta_\alpha, \zeta_z)=(0, 1, 0)$, we find that the corresponding QFI
\begin{equation}
F_{\nu}=\frac{e^{-f_{\alpha}\tau}(\partial_{\nu}f_{\alpha})^2\tau^2(2e^{f_{\alpha}\tau}-1+\cos\theta)\cos^2\frac{\theta}{2}}
{2(e^{f_{\alpha}\tau}-1)}\;.
\end{equation}
Besides, for the scenario where the polarization is parallel to the string we have $(\zeta_r, \zeta_\alpha, \zeta_z)=(0, 0, 1)$, then the corresponding QFI is given by
\begin{equation}
F_{\nu}=\frac{e^{-f_{z}\tau}(\partial_{\nu}f_{z})^2\tau^2(2e^{f_{z}\tau}-1+\cos\theta)\cos^2\frac{\theta}{2}}
{2(e^{f_{z}\tau}-1)}\;.
\end{equation}
Note that for both of the tangential polarization and parallel polarization cases, we find that their corresponding QFIs are also independent of quantum phase $\phi$ of the initial sate. Besides, if the effective distance $r\ll1$, we can find $f_{r}(r,\nu)\approx f_{\alpha}(r,\nu)\approx\frac{3\nu^2(\nu+1)}{\Gamma[2\nu+2]} r^{2(\nu-1)}$, and
$f_{z}(r,\nu)\approx\nu$. Then the QFIs discussed above can be reduced to a  simpler form, and it is interesting to note that in this case the QFI for the
radial and tangential polarization cases is the same.

%%%%%%%%%%%%%%%%%%%%%%%%%%%%%%%%%%%%%%%%%%%%%%%%%%%%%
\begin{figure*}
\centering
\includegraphics[scale=1.2]{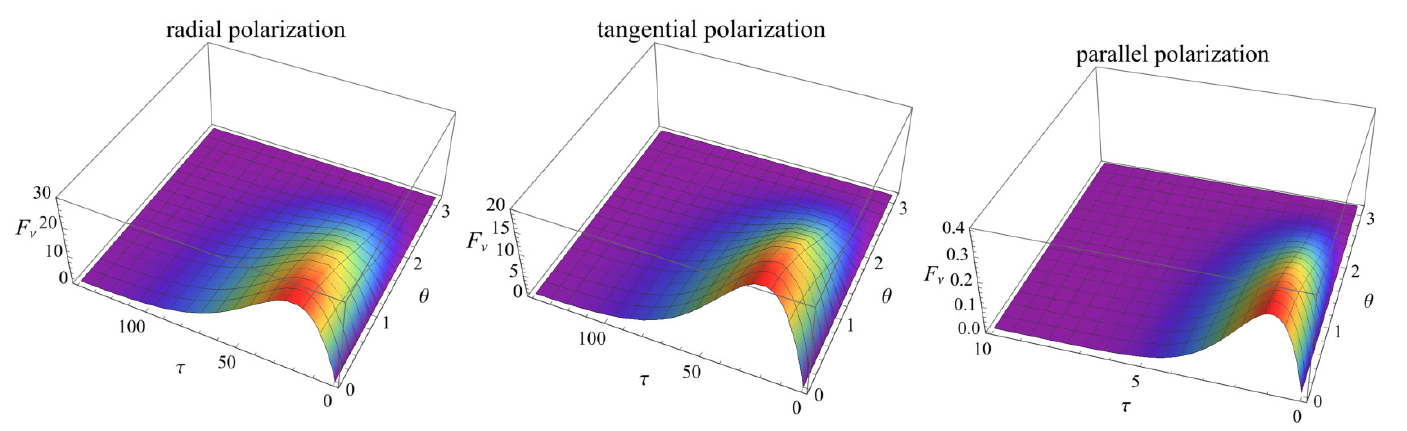}
\caption{(color online). QFI of the deficit angle parameter $\nu$ as a function of the effective time $\tau$ and the initial state parameter $\theta$. The left to right panels correspond to the radial polarization $(\zeta_r, \zeta_\alpha, \zeta_z)=(1, 0, 0)$, the tangential polarization $(\zeta_r, \zeta_\alpha, \zeta_z)=(0, 1, 0)$ and the parallel polarization $(\zeta_r, \zeta_\alpha, \zeta_z)=(0, 0, 1)$, respectively. We take the effective distance $r=0.1$ and the deficit angle parameter $\nu=1.5$. Here $F_{\nu}$ and $\nu$ are dimensionless. $\theta$ is the state parameter, which is expressed in radian. $\tau$ and $r$ are also dimensionless by rescaling, actually $r$ is in the unit of $\frac{c}{\omega_0}$, and $\tau$ is in the unit of $\frac{1}{\gamma_0}$.}\label{fig3}
\end{figure*}

\begin{figure*}
\centering
\includegraphics[scale=1.2]{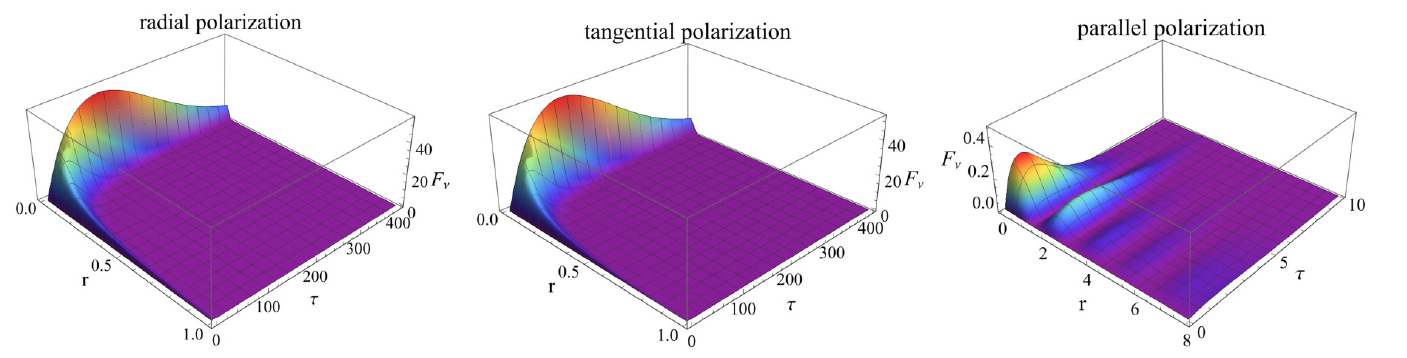}
\caption{(color online). QFI of the deficit angle parameter $\nu$ as a function of the effective time $\tau$ and the effective distance $r$. The left to right panels correspond to the radial polarization $(\zeta_r, \zeta_\alpha, \zeta_z)=(1, 0, 0)$, the tangential polarization $(\zeta_r, \zeta_\alpha, \zeta_z)=(0, 1, 0)$ and the parallel polarization $(\zeta_r, \zeta_\alpha, \zeta_z)=(0, 0, 1)$, respectively. We take the deficit angle parameter $\nu=1.5$ and the initial state parameter $\theta=0$. Here $F_{\nu}$ and $\nu$ are dimensionless. $\theta$ is the state parameter, which is expressed in radian. $\tau$ and $r$ are also dimensionless by rescaling, actually $r$ is in the unit of $\frac{c}{\omega_0}$, and $\tau$ is in the unit of $\frac{1}{\gamma_0}$.}\label{fig4}
\end{figure*}

\begin{figure*}
\centering
\includegraphics[scale=0.52]{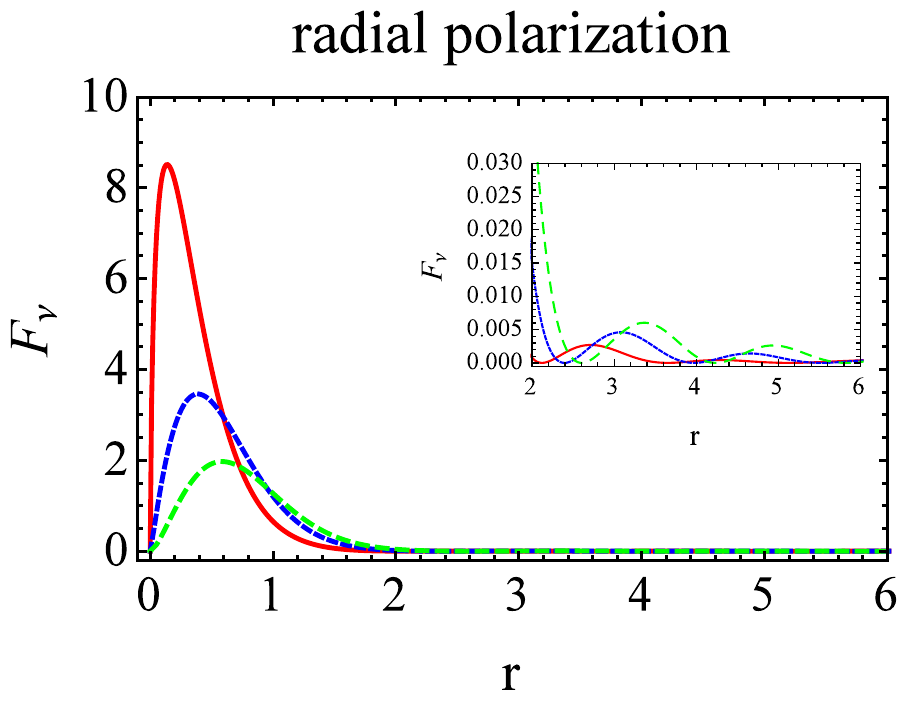}
\includegraphics[scale=0.52]{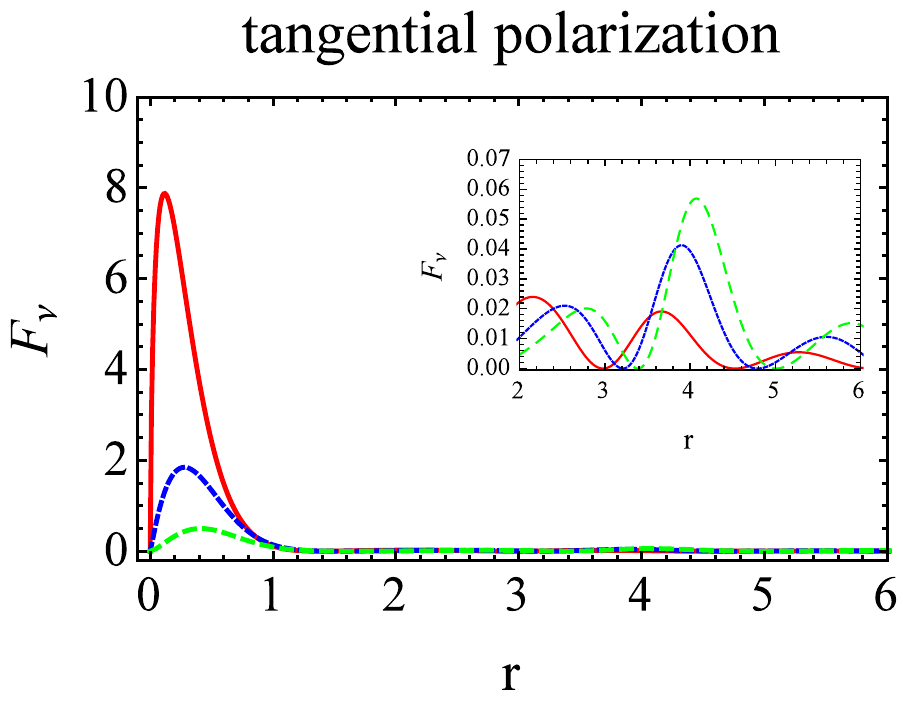}
\includegraphics[scale=0.54]{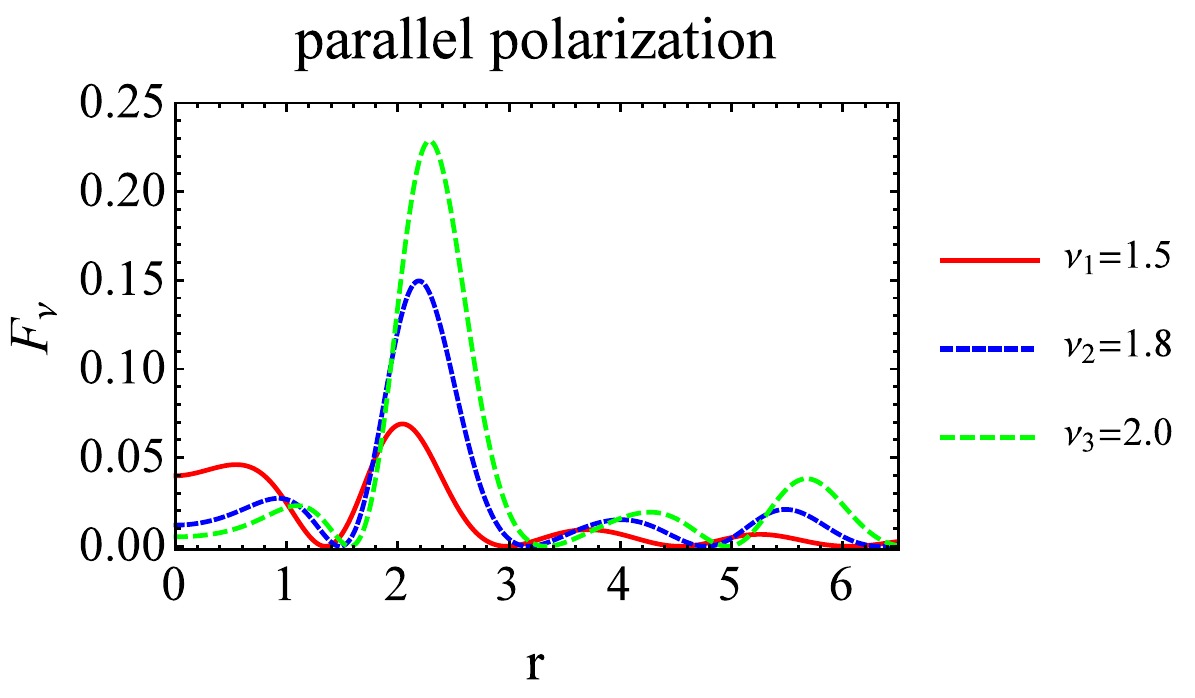}
\caption{(color online). QFI of the deficit angle parameter $\nu$ as a function of the effective distance $r$ with fixed values of the parameter $\nu=1.5, 1.8, 2.0$. The left to right panels correspond to the radial polarization $(\zeta_r, \zeta_\alpha, \zeta_z)=(1, 0, 0)$, the tangential polarization $(\zeta_r, \zeta_\alpha, \zeta_z)=(0, 1, 0)$ and the parallel polarization $(\zeta_r, \zeta_\alpha, \zeta_z)=(0, 0, 1)$, respectively. We take the effective time $\tau=4$ and the initial state parameter $\theta=0$. Here $F_{\nu}$ and $\nu$ are dimensionless. $\theta$ is the state parameter, which is expressed in radian. $\tau$ and $r$ are also dimensionless by rescaling, actually $r$ is in the unit of $\frac{c}{\omega_0}$, and $\tau$ is in the unit of $\frac{1}{\gamma_0}$.}\label{fig5}
\end{figure*}

\subsection{A comparison of the estimation for different polarizations}
In this section, we would like to numerically compare the QFIs of different polarization cases discussed above. To better illustrate our results, we present several figures below.

The QFIs of the deficit angle parameter $\nu$ are plotted as a function of the effective time $\tau$ and the initial states (denoted as different $\theta$), which are shown in Figs.~\ref{fig3}. It is found that the QFIs for different polarization cases behave the same: With the increase of the evolution time, they increase at the beginning to a maximum, then decrease monotonously, and finally vanish. It means that there is an optimal detection time at which
the highest precision for estimating the deficit angle can be possibly achieved during the exploration of cosmic string spacetime. Note that the QFIs for $\theta=\pi$ cases (i.e., the initial detector state is ground state) keep zero over the whole evolution time, which behave quite different from other initial states cases. The reason is that the quantum vacuum fluctuating fields in the cosmic string spacetime can not induce the spontaneous excitation of the static detector, however, can induce its spontaneous emission. Therefore, the detector has no response to the quantum vacuum fluctuating fields when it is initially prepared at the ground state. However, if the detector is initially prepared at the excited state or its quantum superposition with the ground state, its evolution as a consequence of spontaneous emission would depend on the spacetime characteristic through the detector-field interaction. Note that for all the considered polarization cases, the QFIs always achieve their peak values with $\theta=0$ at
a certain time, i.e., the initial excited state. Thus the excited state is the most sensitive state for the detection of cosmic string spacetime. This is because that the interaction between detector and fields, as discussed above, only causes the static atom to emit spontaneously, and thus only the excited state of the detector will response to the vacuum- fluctuation fields as a consequence. We conclude that the maximum sensitivity in the estimation for the deficit angle parameter $\nu$ can be obtained by initial preparation of the detector in the excited state. Note that this conclusion is universal, and it is actually valid for any deficit angle parameter $\nu$ and
the effective distance $r$. To verify its universality, we can see from Eq. \eqref{F2} that when the initial ground state is prepared (i.e., $\theta=\pi$),
the QFI keeps vanishing for any defect angle parameter, evolution time, polarization direction, and effective distance, then we take derivative of  the QFI with respective to
the initial state parameter $\theta$, and find
\begin{equation}\label{Ff2}
\partial_{\theta}F_{\nu}=-\frac{(\partial_{\nu}f(\lambda,r,\nu))^2\tau^2{(e^{f(\lambda,r,\nu)\tau}+\cos\theta)}\sin\theta}
{2e^{f(\lambda,r,\nu)\tau}(e^{f(\lambda,r,\nu)\tau}-1)}.
\end{equation}
Clearly, the solutions to $\partial_{\theta}F_{\nu}=0$ are $\theta=0,\pi$, which means these two values are the points at which the QFI has extreme value.
Since $\partial_{\theta}F_{\nu}\leq0$ for all $\theta\in[0, \pi]$, we obtain $\theta=0$ corresponds to the maximum point and $\theta=\pi$ corresponds to the minimum point. Besides, as shown in the Figure the polarization direction affects the magnitude of the QFI significantly. The maximum QFI magnitude of the radial polarization and the tangential polarization case are almost two orders bigger than the parallel polarization case. We can also find that the response width of the QFI to evolution time behaves differently for different polarizations. The QFI for the parallel polarization case has the narrowest response width compared with the other two cases.

We plot the QFIs of the deficit angle parameter $\nu$ as a function of the effective distance $r$ and the effective time $\tau$ with the deficit angle parameter $\nu=1.5$ shown in Fig.~\ref{fig4}. We find that although different polarizations will cause different behaviors of QFI, for all the polarization cases the optimal position at which the QFIs would achieve their maximum values is the same, i.e., $r\ll1$. It means the optimal precision would be achieved there.  We can also find that the QFIs for the radial polarization and tangential polarization cases are more than two orders larger than that for the parallel polarization case.
Remarkably, when $r\ll1$, the response time for the radial and the tangential polarization cases is much longer than that of parallel polarization.

To further illustrate the dependence of QFI on the position of the detector, by fixing the evolution time we plot the QFI as a function of $r$ with different
fixed deficit angle parameter $\nu$ shown in Fig.~\ref{fig5}. For the radial polarization and the tangential polarization cases, the deficit angle parameter affects the magnitude of the QFI dramatically when $r<1$, and the QFIs keep oscillating while changed relatively slightly when $r>2$. In addition, we can numerically find the
maximum QFIs and the corresponding parameters for all the polarization cases. In our example where three different deficit angle parameters are chosen ($\nu=1.5, 1.8, 2$), we find that for a fixed evolution time and initial state ($\theta=0$) the maximum QFIs for the radial and tangential polarization occur
when $\nu=1.5$ and the detector is closed to the string at $r=0.14$. Correspondingly, the maximum QFIs are $F_{\nu max}=8.513$ for the radial polarization case and $F_{\nu max}=7.796$ for the tangential polarization, respectively. However, unlike the former two cases, the maximum QFI of parallel polarization ($F_{\nu max}=0.2285$) occurs when $\nu=2$ and $r=2.29$. This means that for a fixed time the sensitivity of the detector depends on both the polarization and the deficit angle parameter. Comparing the three maximum QFIs, the former two cases are almost two orders bigger than the latter one obviously.

\begin{figure*}
\centering
\includegraphics[scale=1.2]{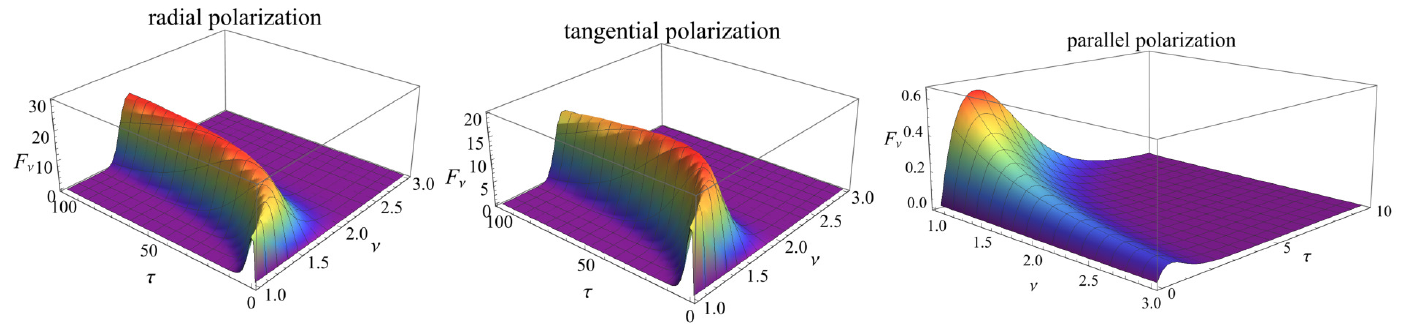}
\caption{(color online). QFI of the deficit angle parameter $\nu$ as a function of itself and the effective time $\tau$ with fixed value of the initial state parameter $\theta=0$. The left to right panels correspond to the radial polarization $(\zeta_r, \zeta_\alpha, \zeta_z)=(1, 0, 0)$, the tangential polarization $(\zeta_r, \zeta_\alpha, \zeta_z)=(0, 1, 0)$ and the parallel polarization $(\zeta_r, \zeta_\alpha, \zeta_z)=(0, 0, 1)$, respectively. We take the effective distance $r=0.1$. Here $F_{\nu}$ and $\nu$ are dimensionless. $\theta$ is the state parameter, which is expressed in radian. $\tau$ and $r$ are also dimensionless by rescaling, actually $r$ is in the unit of $\frac{c}{\omega_0}$, and $\tau$ is in the unit of $\frac{1}{\gamma_0}$.}\label{fig6}
\end{figure*}

Fig.~\ref{fig6} shows how the estimation of the deficit angle $\nu$ is affected by itself.
For the fixed initial states, we find that the QFIs for the radial polarization and tangential polarization cases behave similarly, while they are quite
different from the parallel polarization case. Fixing the time, for the radial and tangential polarization cases, with the increase of the deficit angle $\nu$ the QFIs increase at the beginning, reach a maximum value, then decrease monotonously, and
go to zero eventually. However, for the parallel polarization case, the QFI decreases
monotonously to zero with the increase the deficit angle $\nu$. Again, the QFIs for different polarization cases achieve their maximum values at different deficit angle, and the corresponding maximum values can differ by orders of magnitude.

\section{Discussion}\label{section5}
Recently, many desktop experiments have been designed with atoms and molecules to explore the new physics, and similar model to our scenario has been
used thereof to detect the electromagnetic fields induced by new interaction~\cite{Safronova}. By analyzing the QFI, we in the above have discussed how a two-level detector detects the quantum vacuum fluctuations of electromagnetic fields scattered by the cosmic string spacetime. However, for a more realistic cosmic environment, the thermal noise is inevitable. Consider the relevant analysis in a thermal environment with temperature $T$, the corresponding Wightman function of the electromagnetic fields along the detector's trajectory is given by~\cite{H Cai,huang}
\begin{widetext}
\begin{eqnarray}\label{green2}
G_{rr}(\tau-\tau')&=&\frac{\nu}{8\pi^{2}}\int d\mu_j\csch \frac{\omega_j}{T} \cos(\omega_j\vartriangle\tau+\frac{i\omega_j}{T})
\biggl[\frac{\omega_j}{2}V-\frac{1}{\omega_j}\biggl(\frac{\partial J_{|\nu m|}(k_{\perp}r)}{\partial r}\biggr)^2\biggr]\;,\\
G_{\alpha\alpha}(\tau-\tau')&=&\frac{\nu r^2}{8\pi^{2}}\int d\mu_j\csch \frac{\omega_j}{T} \cos(\omega_j\vartriangle\tau+\frac{i\omega_j}{T}) \biggl[\frac{\omega_j}{2}V-\frac{\nu^2 m^2}{\omega_j r^2}J_{|\nu m|}^2(k_{\perp}r)\biggr]\;,\\
G_{zz}(\tau-\tau')&=&\frac{\nu}{8\pi^{2}}\int d\mu_j\csch \frac{\omega_j}{T} \cos(\omega_j\vartriangle\tau+\frac{i\omega_j}{T})\frac{k_{\perp}^2}{\omega_j}J_{|\nu m|}^2(k_{\perp}r)\;,
\end{eqnarray}
\end{widetext}
with $V=J_{|\nu m+1|}^{2}(k_{\perp}r)+J_{|\nu m-1|}^{2}(k_{\perp}r)$.
Then the associated Fourier transform is
\begin{eqnarray}
{\cal G}(\lambda)=\bigg\{
\begin{array}{c}
\sum_i\frac{e^2|\langle -|r_i|+\rangle|^2\lambda^3}{3\pi\,f^{-1}_{i}(\lambda,r,\nu)}(N(\omega)+1),~~(\omega>0),\\
\sum_i\frac{e^2|\langle -|r_i|+\rangle|^2\lambda^3}{3\pi\,f^{-1}_{i}(\lambda,r,\nu)}N(|\omega|),~~~~~~~~(\omega<0),
\end{array}
\end{eqnarray}
where $N=1/(e^{\omega/T}-1)$ and $f_{i}(\lambda,r,\nu)$ is what we have show in the above. The corresponding coefficients of the Kossakowski matrix $a_{ij}$ is
\begin{eqnarray}\label{A-B2}
\nonumber
A&=&\frac{\gamma_0}{4}\sum_i\zeta_{i}f_{i}(\lambda,r,\nu)(2N+1),
\\
B&=&\frac{\gamma_0}{4}\sum_i\zeta_{i}f_{i}(\lambda,r,\nu).
\end{eqnarray}
For $N=\frac{1}{e^{\omega/T}-1}$, we have let $\hbar=k_{B}=1$. By considering the dimension, we obtain $N=\frac{1}{e^{\hbar\omega/K_{B}T}-1}$. For the reduced Planck constant $\hbar=1.0546\times10^{-34}J\cdot s$, the Boltzmann constant $k_{B}=1.38\times10^{-23}J/K$, the typical transition frequency of the hydrogen atom $\omega\sim10^{15}s^{-1}$~\cite{JHu}, and the temperature for a realistic cosmic environment is $T\cong 2.76 K$~\cite{Turner}, we obtain that $N=\frac{1}{e^{\hbar\omega/k_{B}T}-1}\sim\frac{1}{e^{10^4}}\rightarrow 0$. Then we arrive
\begin{equation}\label{A-B3}
A=B=\frac{\gamma_0}{4}\sum_i\zeta_{i}f_{i}(\lambda,r,\nu)\;.
\end{equation}
It means that one can actually neglect the thermal noise. The corresponding quantum Fisher information is thus
\begin{eqnarray}
\nonumber
F_{\nu}(\omega,\nu,\tau,\theta, r)&=&\frac{e^{-f(\lambda,r,\nu)\gamma_0\tau}(\partial_{\nu}f(\lambda,r,\nu))^2(\gamma_0\tau)^2\cos^2\frac{\theta}{2}}
{2(e^{f(\lambda,r,\nu)\gamma_0\tau}-1)}\\
&&\times(2e^{f(\lambda,r,\nu)\gamma_0\tau}-1+\cos\theta),
\end{eqnarray}
which is the same result to the vacuum case shown above. Therefore, in a more realistic cosmic environment, the results for the vacuum case discussed above are still valid.

Besides, note that actually the cosmic strings have been explored in the analogue gravity with various experimental platforms, e.g., superfluid~\cite{Hendry} and anisotropic medium~\cite{Moraes}, and so on. Our results can also be examined in such analogue gravity systems. When one consider the quantum simulation of cosmic string in a condensed matter system, the density fluctuations thereof can be used to simulate the quantum fields, and an impurity can be introduced as the detector coupled to the density fluctuations. This model is proposed in Refs.~\cite{Recati,Fedichev}, and has been experimentally realized recently~\cite{Schmid,Zipkes}. By controlling the properties of the condensed matter and checking the dynamics of the detector, one can in principle test our results here in the analogue gravity.
\section{conclusion}\label{section6}

Motivated by detecting the topological defects in cosmic string spacetime, we use a static two-level atom which is coupled to the electromagnetic fields as the detector to estimate the deficit angle parameter $\nu$. To obtain optimal estimation conditions, measurements are performed on the detector to make QFI values be maximized for different parameters. We focus on the QFI of $\nu$ with different polarizations, and find that the polarization direction plays a very important role in the estimation of the deficit angle $\nu$, and it not only affects the behavior of the QFI, but also determines its maximum magnitude. For different polarization cases, the QFIs can differ by orders of magnitude. It is found that one can not extract any information about the cosmic string spacetime if the detector is initially prepared at the ground state. The optimal initial state to estimate the deficit angle $\nu$ for all the polarization cases is the pure excited state. The optimal point via the evolution time $\tau$, the effective distance $r$, and the deficit angle $\nu$ itself, at which the QFI achieves the maximum value and thus the possible estimation precision is optimal, depends on the polarization direction. Our results are important extension of quantum metrology to the relativistic framework, and could facilitate the exploration of cosmic string spacetime characteristic in the future.

\begin{acknowledgments}
Z. T. was supported by the National Natural Science Foundation of China under Grant No. 11905218, and the CAS Key Laboratory for Research in Galaxies and Cosmology, Chinese Academy of Sciences (No. 18010203). Y. Y. was supported by the National Natural Science Foundation of China under Grant No. 12105097, and Scientific Research Fund of Hunan Provincial Education Department(No. 20C0787). J. J. was supported by the National Natural Science Foundation of China under Grant No. 11875025.
\end{acknowledgments}


\begin{thebibliography}{00}
\bibitem{Kaiser}
N. Kaiser and A. Stebbins, Microwave anisotropy due to cosmic strings, Nature {\bf 310}(5976), 391-393 (1984).
\bibitem{Bennett}
D. P. Bennett, and F. R. Bouchet, Cosmic-string evolution,
Phys. Rev. Lett. {\bf63}(26), 2776 (1989).
\bibitem{Vilenkin}
A. Vilenkin and E. P. S. Shellard, Cosmic Strings and Other Topological Defects (Cambridge
University Press, Cambridge, 1994).
\bibitem{Hindmarsh}
M. B. Hindmarsh and T. W. B. Kibble, Cosmic strings, Rep.
Prog. Phys. {\bf58}(5), 411 (1995).
\bibitem{Brandenberger1}
R. H. Brandenberger, A. T. Sornborger, and M. Trodden, $\gamma$-ray bursts from ordinary cosmic strings, Phys. Rev. D {\bf48}(2), 940 (1993).
\bibitem{MacGibbon}
J. H. MacGibbon and R. H. Brandenberger, $\gamma$-ray signatures from ordinary cosmic strings, Phys. Rev. D {\bf47}(6), 2283 (1993).
\bibitem{Damour}
T. Damour and A. Vilenkin, Gravitational radiation from cosmic (super)strings: Bursts, stochastic background, and observational windows,
Phys. Rev. D {\bf71}(6), 063510 (2005).
\bibitem{Brandenberger2}
R. Brandenberger, H. Firouzjahi, J. Karouby, and S. Khosravi, Gravitational radiation by cosmic strings in a junction, J. Cosmol. Astropart. Phys. {\bf2009}(01), 008 (2009).
\bibitem{Jackson}
M. G. Jackson and X. Siemens, Gravitational wave bursts from cosmic superstring reconnections,
J. High Energy Phys. {\bf2009}(06), 089 (2009).
\bibitem{Cheng}
K. S. Cheng, Y. Yu, and T. Harko, High-redshift gamma-ray bursts: observational signatures of superconducting cosmic strings?, Phys. Rev. Lett. {\bf104}(24), 241102 (2010).
\bibitem{Kibble}
T. W. B. Kibble, Topology of cosmic domains and strings, J. Phys. A {\bf9}(8), 1387 (1976).
\bibitem{Davis1}
A. C. Davis and T. W. B. Kibble, Fundamental cosmic strings, Contemporary Physics, {\bf46}(5), 313-322 (2005).
\bibitem{Siemens}
S. Xavier, M. Vuk, and C. Jolien, Gravitational-wave stochastic background from cosmic strings, Phys. Rev. Lett. {\bf98}(11), 111101 (2007).
\bibitem{Linet}
B. Linet, Quantum field theory in the space-time of a cosmic string,
Phys. Rev. D {\bf35}(2), 536 (1987).
\bibitem{Frolov}
V. P. Frolov and E. M. Serebriany, Vacuum polarization in the gravitational field of a cosmic string,
Phys. Rev. D {\bf35}(12), 3779 (1987).
\bibitem{Davies1}
P. C. Davies and V. Sahni, Quantum gravitational effects near cosmic strings,
Class. Quantum Grav. {\bf5}(1), 1 (1988).
\bibitem{Gott}
J. R. Gott, Gravitational lensing effects of vacuum strings-Exact solutions, Astrophys. J. {\bf288}, 422 (1985).
\bibitem{Charnock}
T. Charnock, A. Avgoustidis, E. Copeland, and M. A., CMB constraints on cosmic strings and superstrings,
Phys. Rev. D {\bf93}(12), 123503 (2016).
\bibitem{Foreman}
S. Foreman, M. Adam, and S. Douglas, Predicted constraints on cosmic string tension from Planck and future CMB polarization measurements, Phys. Rev. D {\bf84}, 043522 (2011).
\bibitem{Dvorkin}
C. Dvorkin, W. Mark, and H. Wayne, Cosmic string constraints from WMAP and the South Pole Telescope data, Phys. Rev. D {\bf84}(12), 123519 (2011).
\bibitem{Eiichiro}
C. L.Bennett, R. S. Hill, G. Hinshaw, et al. Seven-year wilkinson microwave anisotropy probe (WMAP*) observations: Are there cosmic microwave background anomalies?, The Astrophysical Journal Supplement Series {\bf192}(2), 18 (2011).

\bibitem{Aghanim}
N. Aghanim, et al. Planck 2015 results-XI. CMB power spectra, likelihoods, and robustness of parameters,
Astronomy Astrophysics {\bf594}, A11 (2016).
\bibitem{Urrestilla}
J. Urrestilla, et al. Cosmic string parameter constraints and model analysis using small scale Cosmic Microwave Background data. J. Cosmol. Astropart. Phys. {\bf2011}(12), 021 (2011).
\bibitem{Helliwell}
T. M. Helliwell and D. A. Konkowski, Vacuum fluctuations outside cosmic strings,
Phys. Rev. D {\bf34}(6), 1918 (1986).

\bibitem{Sousa}
P. de Sousa Gerbert and R. Jackiw, Classical and quantum scattering on a spinning cone, Commun. Math. Phys. {\bf124}(2), 229 (1989).
\bibitem{Bezerra}
V. B. Bezerra, Gravitational Aharonov-Bohm effect in a locally flat spacetime, Class. Quantum Grav. {\bf8}(10), 1939 (1991).

\bibitem{Iliadakis}
L. Iliadakis, U. Jasper, and J. Audretsch, Quantum optics in static spacetimes: How to sense a cosmic string,
Phys. Rev. D {\bf51}(6), 2591 (1995).
\bibitem{Berezinsky}
V. Berezinsky, B. Hnatyk, and A. Vilenkin, Gamma ray bursts from superconducting cosmic strings, Phys. Rev. D {\bf64}(4), 043004 (2001).
\bibitem{Davies}
P. C. Davies and V. Sahni, Quantum gravitational effects near cosmic strings, Class. Quantum Grav. {\bf5}(1), 1 (1988).
\bibitem{Bilge}
A. H. Bilge, M. Hortacsu, and N. Ozdemir, Can an Unruh detector feel a cosmic string?, Gen. Relativ. Grav. {\bf30}(6), 861 (1998).

\bibitem{Giovannetti}V. Giovannetti, S. Lloyd, and L. Maccone, Quantum-enhanced measurements: beating the standard quantum limit, Science {\bf306}(5700), 1330 (2004).
\bibitem{Giovanetti2} V. Giovannetti, S. Lloyd, and L. Maccone, Quantum Metrology,
Phys. Rev. Lett. {\bf96}(1), 010401 (2006).
\bibitem{Paris}
M. G. A. Paris, Quantum estimation for quantum technology, International Journal of Quantum Information, {\bf7}(supp01), 125 (2009).
\bibitem{advance}
V. Giovannetti, S. Lloyd, and L. Maccone, Advances in quantum metrology, Nature Photonics {\bf5}(4), 222 (2011).

\bibitem{Ahmadi}
A. Mehdi, E. B. David, and F. Ivette, Quantum metrology for relativistic quantum fields, Phys. Rev. D {\bf89}(6), 065028 (2014).
\bibitem{Huang}
X. Huang, J. Feng, Y. Z. Zhang, and H. Fan, Quantum estimation in an expanding spacetime, Annals of Physics, {\bf397}, 336 (2018).
\bibitem{Kish}
S. P. Kish and T. C. Ralph, Quantum metrology in the Kerr metric, Phys. Rev. D {\bf99}(12), 124015 (2019).

\bibitem{xiaobao}
X. Liu, J. Jing, J. Wang \emph{et al.}. Optimal estimation of parameters for scalar field in an expanding spacetime exhibiting Lorentz invariance violation.
Quantum Inf Process {\bf 19}(1), 26 (2020).

\bibitem{Mann R B}
H. Du, R. B. Mann, Fisher information as a probe of spacetime structure: relativistic quantum metrology in (A) dS, J. High Energy Phys. {\bf2021}(5), 1-22 (2021).
\bibitem{Parameter estimation}
D. Hosler, P. Kok, Parameter estimation using NOON states over a relativistic quantum channel, Phys. Rev. A  {\bf88}(5), 052112 (2013).
\bibitem{relativistic motion}
C. Y. Huang, W. Ma, D. Wang,\emph{et al.} How the relativistic motion affect quantum Fisher information and Bell non-locality for multipartite state, Scientific reports {\bf7}(1), 1-8 (2017).
\bibitem{Liu1}
X. Liu, Z. Tian, J. Wang, and J. Jing, Relativistic motion enhanced quantum estimation of $\kappa $ -deformation of spacetime, The European Physical Journal C {\bf78}(8), 1-9 (2018).
\bibitem{Liu2}
X. Liu, J. Jing, Z.  Tian, and W. Yao, Does relativistic motion always degrade quantum Fisher information? Phys. Rev. D {\bf103}(12), 125025 (2021).

\bibitem{M-Ahmadi}
M. Ahmadi, D. E. Bruschi, C. Sab\'in, G. Adesso, and I. Fuentes, Scientific Reports {\bf 4}, 4996 (2014).

\bibitem{Tian}
Z. Tian, J. Wang, H. Fan, and J. Jing, Relativistic quantum metrology in open system dynamics, Scientific reports {\bf5}(1), 1-6 (2015).
\bibitem{Adesso}
M. Aspachs, G. Adesso, and I. Fuentes, Optimal Quantum Estimation of the Unruh-Hawking Effect,
Phys. Rev. Lett. {\bf 105}(15), 151301 (2010).

\bibitem{TianZH}
Z. Tian, J. Wang, J. Jing, and A. Dragan, Entanglement enhanced thermometry in the detection of the Unruh effect, Annals of Physics {\bf377}, 1-9 (2017).

\bibitem{gravitational wave}
R. Schnabel, N. Mavalvala, D. McClelland, \emph{et al.}, Quantum metrology for gravitational wave astronomy, Nat. Commun. {\bf1}(1), 1-10 (2010).
\bibitem{Schwarzschild spacetime parameters}
D. E. Bruschi, A. Datta, R. Ursin, et al., Quantum estimation of the Schwarzschild spacetime parameters of the Earth, Phys. Rev. D {\bf90}(12), 124001 (2014).
\bibitem{gravitational redshift}
H. M\"{u}ller, A. Peters, S. Chu, A precision measurement of the gravitational redshift by the interference of matter waves, Nature {\bf463}(7283), 926 (2010).

\bibitem{Helstrom}
C. W. Helstrom, Quantum Detection and Estimation Theory, Journal of Statistical Physics {\bf1}(2), 231-252 (1969).
\bibitem{Holevo} A. S. Holevo, Probabilistic and statistical aspects of quantum theory[M]. (Edizioni della Normale, Pisa, 1982).

\bibitem{jin}
Y. Jin, Precision protection through cosmic string in quantum metrology, The European Physical Journal C {\bf80}(12), 1-7 (2020).
\bibitem{huang}
Z. Huang, H. Situ, Z. He, Quantum Fisher information in the cosmic string spacetime, Class. Quant. Grav. {\bf37}(17), 175002 (2020).
\bibitem{conicity}
W. Cong, et al. Quantum detection of conicity, Phys. Let. B {\bf820}, 136482 (2021).
\bibitem{ying}
Y. Yang, J. Wang, M. Wang, J. Jing and Z. Tian, Parameter estimation in cosmic string space-time by using the inertial and accelerated detectors, Class. Quant. Grav. {\bf37}(6), 065017 (2020).
\bibitem{Cramer}
H. Cram\'{e}r, Mathematical Methods of Statistics (PMS-9), Volume 9[M]. (Princeton University, Princeton, 2016).
\bibitem{Zhong}
W. Zhong, Z. Sun, J. Ma, X. Wang, and F. Nori, Fisher information under decoherence in Bloch representation, Phys. Rev. A {\bf87}(2), 022337 (2013).
\bibitem{Skarzhinsky}
V. D. Skarzhinsky, D. D. Harari, and U. Jasper, Quantum electrodynamics in the gravitational field of a cosmic string, Phys. Rev. D {\bf49}(2), 755 (1994).
\bibitem{Pavel}
K. Pavel, Electromagnetic field near a cosmic string, Phys. Rev. D {\bf74}(6), 065006 (2006).
\bibitem{Heinz}
H. Breuer and F. Petruccione, The Theory of Open Quantum Systems[M]. (Oxford University, Oxford, 2002).
\bibitem{Lindblad}
G. Lindblad, On the generators of quantum dynamical semigroups, Commun. Math. Phys. {\bf 48}(2), 119 (1976).
\bibitem{Kimura}
G. Kimura, The Bloch vector for N-level systems, Phys. Lett. A {\bf314}(5-6), 339 (2003).

\bibitem{Safronova}
M. S. Safronova, D. Budker, D. DeMille, D. F. J. Kimball, A. Derevianko, and C. W. Clark, Search for new physics with atoms and molecules,
Rev. Mod. Phys. {\bf 90}, 025008 (2018).

\bibitem{H Cai}
H. Cai, H. Yu, and W. Zhou, Spontaneous excitation of a static atom in a thermal bath in cosmic string spacetime, Phys. Rev. D {\bf92}(8), 084062 (2015).
\bibitem{JHu}
J. Hu and H. Yu, Geometric phase for an accelerated two-level atom and the Unruh effect, Phys. Rev. A {\bf85}(3), 032105 (2012).
\bibitem{Turner}
M. S. Turner, Why is the temperature of the universe 2.726 Kelvin, Science, {\bf262}(5135), 861-867(1993).
\bibitem{Hendry}
P. C. Hendry, N. S. Lawson, R. A. M. Lee, et al. Generation of defects in superfluid 4He as an analogue of the formation of cosmic strings, Nature {\bf368}(6469), 315-317 (1994).
\bibitem{Moraes}
C. Satiro and F. Moraes,  A liquid crystal analogue of the cosmic string, Modern Physics Letters A, {\bf20}(33), 2561-2565 (2005).
\bibitem{Recati}
A. Recati, P. O. Fedichev, W. Zwerger, J. von Delft, and P. Zoller, Atomic quantum dots coupled to a reservoir of a superfluid Bose-Einstein condensate, Phys. Rev. Lett. {\bf94}(4), 040404(2005).
\bibitem{Fedichev}
P. O. Fedichev and U. R. Fischer, Gibbons-Hawking effect in the sonic de Sitter space-time of an expanding Bose-Einstein-condensed gas, Phys. Rev. Lett. {\bf91}(24), 240407 (2003).
\bibitem{Schmid}
S. Schmid, A. Harter, and J. H. Denschlag, Dynamics of a cold trapped ion in a Bose-Einstein condensate, Phys. Rev. Lett. {\bf105}(13), 133202 (2010).
\bibitem{Zipkes}
C. Zipkes, S. Palzer, C. Sias, and M. Kohl, A trapped single ion inside a BosešCEinstein condensate, Nature {\bf464}(7287), 388-391 (2010).

\end{thebibliography}
\end{document}